\documentclass[conference,10pt]{IEEEtran}
\usepackage{epsfig,rotating,setspace,latexsym,amsmath,epsf,amssymb,amsfonts,bm,subfigure,epstopdf,cite,authblk,bbm,mathtools,amsthm,algorithm,color,systeme,mathtools,amsmath,derivative,bigints,cite,amsmath,amssymb,amsfonts, amsthm,algorithmic,graphicx,textcomp,xcolor,bbm}

\usepackage[
  paperwidth=8.5in,
  paperheight=11in,
  left=0.673in,
  right=0.660in,
  top=0.72in,
  bottom=1.013in
]{geometry}

\def\BibTeX{{\rm B\kern-.05em{\sc i\kern-.025em b}\kern-.08em
    T\kern-.1667em\lower.7ex\hbox{E}\kern-.125emX}}

\newtheorem{remark}{Remark}
\newtheorem{corollary}{Corollary}
\newtheorem{definition}{Definition}
\newtheorem{theorem}{Theorem}

\newtheorem{prop}{Proposition}

\IEEEoverridecommandlockouts

\begin{document}

\title{Age of Job Completion Minimization with \\ Stable Queues}
\author{Stavros Mitrolaris \qquad Subhankar Banerjee \qquad Sennur Ulukus\\
\normalsize Department of Electrical and Computer Engineering\\
\normalsize University of Maryland, College Park, MD 20742\\
\normalsize \emph{stavros@umd.edu} \qquad \emph{sbanerje@umd.edu} \qquad \emph{ulukus@umd.edu}}

\maketitle

\begin{abstract}
We consider a time-slotted job-assignment system with a central server, $N$ users and a machine which changes its state according to a Markov chain (hence called a \emph{Markov machine}). The users submit their jobs to the central server according to a stochastic job arrival process. For each user, the server has a dedicated job queue. Upon receiving a job from a user, the server stores that job in the corresponding queue. When the machine is not working on a job assigned by the server, the machine can be either in \textit{internally busy} or in \textit{free} state, and the dynamics of these states follow a binary symmetric Markov chain. Upon sampling the state information of the machine, if the server identifies that the machine is in the free state, it schedules a user and submits a job to the machine from the job queue of the scheduled user. To maximize the number of jobs completed per unit time, we introduce a new metric, referred to as the \textit{age of job completion}. To minimize the age of job completion and the sampling cost, we propose two policies and numerically evaluate their performance. For both of these policies, we find sufficient conditions under which the job queues will remain stable.
\end{abstract}

\section{Introduction}
In mission-critical control applications such as intelligence and surveillance, job offloading to edge computing devices is a fundamental operation. In such applications, an edge device is often shared among multiple users or servers, each independently generating computational tasks. The job offloading process typically occurs in a stochastic manner, leading to random and competing task arrivals at the edge. Thus, when a user attempts to offload a job to the edge computing device, the availability of the edge computing device becomes highly uncertain. Therefore, \textit{efficient tracking} of the operational state of the devices and \textit{timely offloading} of computational tasks are essential to ensure effective system performance.

Due to the randomness in the state information of the edge computing device, we can model it as a stochastic process. The problem of tracking or remotely estimating such stochastic processes has been extensively studied in the literature. In particular, significant attention has been given to the remote estimation of Markov processes. Numerous studies have explored various aspects of this problem, including optimal sampling, estimation policies, and scheduling strategies, e.g.. see \cite{maatouk2020age, kam2020age, chen2021scheduling, kriouile2022minimizing}, and the references therein. In all these works, the sampling process does not change the dynamics of the underlying Markov chain. Thus, it is important to note that the aforementioned studies do not consider the presence of an external control that can actively influence the state evolution of the system. However, in edge computing scenarios, each job offloading event may alter the state of the device, thereby introducing a control-dependent dynamic. Motivated by this, in the literature, recent studies investigate the problem of tracking a Markov process under external control inputs \cite{banerjee2025tracking, liyanaarachchi2025optimum, chamoun2025edge, liyanaarachchi2025age, sariisik2025maximize, chamoun2025mappo}. These studies are mainly done in the context of job offloading to computing devices whose dynamics can be modeled as a Markov process. Consequently, the term \textit{Markov machine} has been introduced in the literature to describe such devices.

\begin{figure}[t]
    \centerline{\includegraphics[width = 0.894\columnwidth]{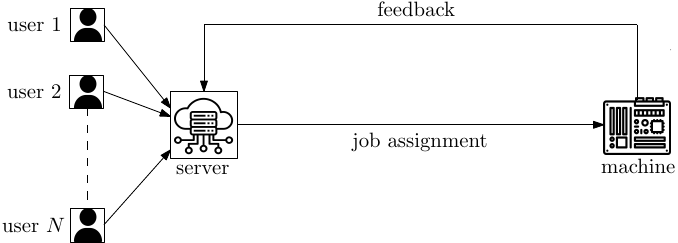}}
    \caption{A job-assignment system with a Markov machine, a central server and $N$ users.}
    \label{fig:1}
    \vspace*{-0.4cm}
\end{figure}

In this work, we consider a job-assignment system, where $N$ users present in the system stochastically submit their job requests to a central server, as shown in Fig.~\ref{fig:1}\footnote{The icons are taken from www.flaticon.com.}. For each user, the central server has a job queue, and it stores the job requests of a user in its dedicated job queue. The system includes a Markov machine that serves a job request assigned by the central server. At the beginning of a time slot, the server may sample the state information of the machine, paying a sampling cost, and if it observes that the machine is in a free state, it schedules a user and submits a job from the corresponding job queue to the machine. As we are considering a job-offloading system, our main goal is to maximize the number of jobs completed per unit time, and to this end, we introduce a novel metric termed the \textit{age of job completion}. As will be discussed later, neither this metric nor a similar metric has been considered in the existing literature on job assignment to Markov machines. To minimize the age of job completion and the sampling cost, we propose two policy pairs, each comprising a scheduling policy and a sampling policy. For both policy pairs, we numerically find their performance. Furthermore, we derive sufficient conditions that ensure the stability of the job queues for both policy pairs. To the best of our knowledge, such stability results have not been explored in prior studies on job offloading to Markov machines.  

Due to the space limitation, we omit the proofs of our results here, which we will provide in a journal version on arXiv.

\section{Related Works}
In \cite{banerjee2025tracking}, the authors considered a Markov machine equipped with a sampler that samples and transmits its state information to a central server at the cost of sampling. To quantify the tracking performance, the authors introduced a tailored version of the age of incorrect information (AoII) metric. The AoII metric does not directly address the maximization of the number of completed jobs, whereas the metric proposed here, i.e., the age of job completion metric, better captures this objective in job-offloading systems.

In \cite{liyanaarachchi2025optimum}, the authors considered a network of Markov machines, where the jobs arrive at a resource allocator following a Poisson process. The authors studied an optimal sampling rate allocation problem for query-based sampling of the machines. To monitor the tracking performance, the authors used the well-known \textit{binary freshness} (BF) metric. They also introduced the \textit{false rejection ratio} (FRR) and the \textit{false acceptance ratio} (FAR) metrics to quantify the effects of limited tracking: the former measures the fraction of jobs rejected despite the presence of a free machine due to insufficient tracking, while the latter measures the fraction of jobs accepted even though no machine was actually free, again due to limited tracking. It is important to note that, similar to AoII, none of these metrics directly captures the objective of maximizing the number of completed jobs. Moreover, the authors in \cite{liyanaarachchi2025optimum} did not consider a job queue at the resource allocator. In \cite{sariisik2025maximize}, the authors considered a job assignment system with a network of exhausted workers and a resource allocator, where the efficiency of each worker depends on its current state. Similar to \cite{liyanaarachchi2025optimum}, the authors formulated and studied an optimal sampling rate allocation problem for query-based sampling of these workers, aiming to maximize the probability that each worker operates in its most efficient state.

A revenue maximization problem has been considered in a job assignment system in \cite{liyanaarachchi2025age}. They considered that a resource allocator has a unit-size buffer, i.e., it can store at most one job at a time. In \cite{chamoun2025mappo}, the authors considered a job-dispatching system with a network of dispatchers and edge servers; these servers were modeled as Markov machines. Their goal was to maximize the differential gain between job completion and sampling cost, and they formulated the problem as a discounted Markov decision process (MDP). As a result, their work did not directly maximize the average number of jobs completed over the time horizon. When a job was dispatched to a server, it was stored in a limited-size job queue, and if the queue was full, the oldest job was replaced. Similar to \cite{chamoun2025mappo}, the authors of \cite{chamoun2025edge} considered a job-dispatching system for a network of dispatchers and edge servers; however, in contrast to \cite{chamoun2025mappo}, they did not consider a job queue at the edge server, i.e., if a dispatcher offloads a job to an edge server and if it is busy, the job is dropped. They proposed an index-based policy that maximizes the probability that a dispatched job is accepted by an edge server. However, in our work, the central server first samples the state of the machine and submits a job only if the machine is free, resulting in an acceptance probability of $1$ for offloaded jobs. Thus, our focus is on maximizing the number of jobs completed per unit time, rather than the job acceptance probability as in \cite{chamoun2025edge}. 

In contrast to \cite{banerjee2025tracking, liyanaarachchi2025optimum, chamoun2025edge, liyanaarachchi2025age, sariisik2025maximize, chamoun2025mappo}, in this work, we consider limitless full job queues, making the study of queue stability a necessary aspect of our problem. 

\section{System Model and Problem Formulation}
In our job assignment system, user $i$ submits a job request to the central server at the end of each time slot, according to an i.i.d.~Bernoulli random variable $a_{i}(t)$, with probability of success being $p_{i}>0$. When the machine is not executing a job assigned by the server, denoted as an \textit{external job}, the machine can be either free or busy processing an internal job. We model the dynamics between these free and \textit{internally busy} states with a binary symmetric Markov chain, with probability of state transition being $q>0$.\footnote{For simplicity of computations, in this work, we consider a symmetric state transition, and the general case will be considered in a journal version.} 

The machine only accepts an external job when it is in the free state. Thus, when the machine is not executing an external job and the server intends to assign a job to the machine, the server samples the state information of the machine, which incurs a cost $L$, and receives the sampled state information instantaneously. We denote this sampling operation at time $t$ with an indicator variable $\mu(t)$, where $\mu(t)=1$ implies that the server samples the state information of the machine at time $t$, and $\mu(t)=0$ implies otherwise. If the server infers that the machine is in a free state, the server instantaneously assigns a job to the machine from one of the non-empty job queues. At time $t$, we denote this scheduling operation with a vector $(\pi_{i}(t))_{i=1}^{N}$. At time $t$, if the server schedules a job from the job queue of user $i$, then $\pi_{i}(t)=1$; otherwise $\pi_{i}(t)=0$. Thus, we represent an employable policy by the central server as 
\begin{align}
    \phi=(\mu(t), (\pi_{i}(t))_{i=1}^{N})_{t=1}^{\infty}.
\end{align} 

We assume that the server performs the sampling, scheduling and job assignment processes, at the beginning of a time slot and the machine completes either an internal job or an external job at the end of a time slot. When the machine is processing a job of user $i$, the service time follows a geometric distribution with parameter $q_i >0$. During this period, the machine cannot process any other job, and the server is aware of the state of the machine. Once the job is completed, the server is instantaneously notified. After the machine completes a job, it transits to the internally busy state with probability $s>0$, or to the free state with probability $(1-s)$. Thus, if an external job is completed at time slot $t-1$, the server becomes uncertain about the state of the machine at the beginning of slot $t$ and must sample its state information before submitting a future job. 

\begin{figure}[t]
    \centerline{\includegraphics[width = 0.95\columnwidth]{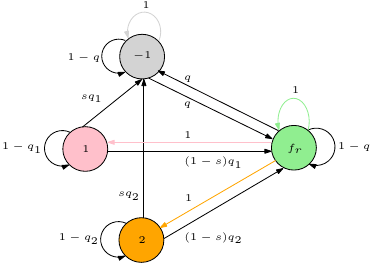}}
    \caption{In this figure, we consider a system with $N=2$ users, and we show the corresponding state transitions of the Markov machine. The transitions labeled with black arrows represent slot-by-slot transitions, whereas those labeled with green, grey, orange, and pink arrows denote instantaneous transitions resulting from the server actions, namely scheduling and job assignment. When the machine is in the free state while both job queues are empty, and the server samples the state, the machine remains free, and we denote this self-transition with the green arrow. Similarly, when the machine is in the free state, and the server samples and assigns a job from the job-queue of the first (second) user to the machine, the machine instantaneously transitions to state $1$ ($2$), denoted by the pink (orange) arrow. }
    \label{fig:2}
    \vspace*{-0.4cm}
\end{figure}

At time $t$, we denote the last known state information of the machine at the server with $x(t) \in \{-1, 0, \ldots, N, f_{r}\}$.
Here $-1$ corresponds to the internally busy state, the state $0$ denotes the ambiguity of the state of the machine after it completes a job, $i \in \{1, \ldots, N \}$ indicates that the machine is processing a job of user $i$,   and $f_{r}$ denotes that the machine is free. Note that, $x(t)$ gets updated if the server samples the state of the machine at time $t$, or receives a job completion acknowledgment at $t-1$, otherwise $x(t)$ evolves as $x(t)=x(t-1)$. Note that state $f_r$ is only achieved when the sampled state of the machine is free and the server does not schedule any job. Furthermore, $x(t)$ remains $0$ from the time a job is completed until the server next samples the state information. Similarly, with $x_{a}(t)$, we denote the actual state of the machine at time $t$. The variable $x_{a}(t)$ takes the values $\{-1,1,2\cdots,N,f_{r}\}$, retaining the same interpretation as $x(t)$. Fig.~\ref{fig:2} shows the dynamics of the considered Markov machine.

\begin{figure}[t]
    \centerline{\includegraphics[width = 0.85\columnwidth]{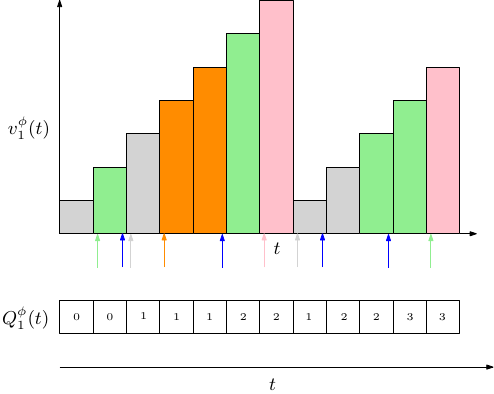}}
    \vspace*{-0.2cm}
    \caption{We consider a system with $N=2$ users. The figure illustrates a sample path of age of job completion and the corresponding job queue length for the first user under an arbitrary policy $\phi$. At each time, we color the age block based on the current state of the machine. We note that a job arrives at the end of a time slot which we denote with blue arrows, and the server performs sampling and job assignment at the beginning of a slot with  colored arrows following the same color conventions as in Fig.~\ref{fig:1}. For example, during time slot $3$, the machine is internally busy and transitions to the free state at the end of the slot. At the beginning of slot $4$, the server samples the state of the machine and  schedules a job from the job queue of user $2$ instantaneously.}
    \label{fig:3}
    \vspace*{-0.4cm}
\end{figure}

At time $t$, for a policy $\phi$, we denote the number of jobs at the queue corresponding to the $i$th user with $Q_{i}^{\phi}(t)$. For user $i$, we denote the job completion event with an indicator variable $b_{i}^{\phi}(t)$, where $b_{i}^{\phi}(t)=1$ implies a successful job completion, whereas $b_{i}^{\phi}(t)=0$ implies otherwise. We describe the queue dynamics as, 
\begin{align}
     Q_{i}^{\phi}(t+1) = \max \{ Q_{i}^{\phi}(t) - b_{i}^{\phi}(t), 0 \} + a_i(t).
\end{align}
To account for the importance of timely job completions, we introduce the metric of \emph{age of job completion}. For user $i$ at time $t$, we measure the corresponding age of job completion as the time since the last completion of its job, and we denote it with $v_{i}^{\phi}(t)$, i.e.,  $v_{i}^{\phi}(t) = t- \sup\{{t': t'<t, \  b_{i}^{\phi}(t')=1}\}$, see Fig.~\ref{fig:3}. We denote the average age of job completion of user $i$, with $\Delta_{i}^{\phi}$ and the total age with $\Delta^{\phi}$, i.e., 
\begin{align}
    \Delta_{i}^{\phi} =& \limsup_{T\rightarrow \infty} \frac{1}{T} \sum_{t=1}^{T} \mathbb{E}_{\phi}[v_{i}^{\phi}(t)], \\ 
    \Delta^{\phi} = &\frac{1}{N}{\sum_{i=1}^{N}\Delta_{i}^{\phi}}.\label{eq:2}
\end{align}

We denote the average sampling cost with $S^{\phi}$, i.e.,
\begin{align}
    S^{\phi} = \limsup_{T\rightarrow\infty} \frac{L}{T} \sum_{t=1}^{T} \mathbb{E}_{\phi}[\mu(t)].
\end{align}
Thus, the total average cost, i.e., the age of job completion together with the sampling cost, is
\begin{align}
    \Delta^{\phi} + S^{\phi}.
\end{align}
For simplicity, we omit the explicit dependence of the defined variables on $\phi$ whenever obvious from the context. Note that job requests are not dropped, but instead remain in the queue until they are served. Thus, we study sampling policies that minimize the total cost while maintaining stable job queues.

Intuitively, a policy $\phi$ that minimizes (\ref{eq:2}) also maximizes the number of  completed jobs per unit time. This argument is formally shown in a subsequent work. In this work, first we present an adaptive randomized policy that comprises an \textit{adaptive randomized sampling policy} and an \textit{adaptive randomized scheduling policy}. Then, we study the max-age scheduling policy together with an \textit{adaptive randomized sampling policy}.

At time $t$, we observe the non-empty job queues and schedule the corresponding users with a stationary randomized policy. Whenever an identical set of job queues becomes non-empty, the same stationary randomized policy is applied. We call this policy an adaptive randomized scheduling policy. Similarly, based on the non-empty job queues, we define an adaptive randomized sampling policy. We provide a formal definition of an adaptive randomized policy in Definition~\ref{def:1}.

\begin{definition} \label{def:1}
    An adaptive randomized policy is specified by the collection
    \begin{align}
        \Pi = \{\, (\mu(\mathcal{S}), \bm{\pi}(\mathcal{S}) ) : \mathcal{S} \subseteq [N],\, \mathcal{S} \neq \varnothing \,\},
    \end{align}
    where  for each non-empty subset $\mathcal{S} \subseteq [N] := \{1,2,\ldots,N\}$, $\mu(\mathcal{S}) \in (0,1]$ is the sampling probability and
    \begin{align*}
        \bm{\pi}(\mathcal{S}) = (\pi_1(\mathcal{S}), \pi_2(\mathcal{S}), \ldots, \pi_N(\mathcal{S}))
    \end{align*}
    is a probability distribution over users  satisfying
    \begin{align}
        \pi_i(\mathcal{S}) > 0 \text{ if } i \in \mathcal{S}, 
        \
        \pi_i(\mathcal{S}) = 0 \text{ if } i \notin \mathcal{S}, 
        \
        \sum_{i \in \mathcal{S}} \pi_i(\mathcal{S}) \!=\! 1.
    \end{align}
    At each time slot $t$, let $\mathcal{S}(t) = \{ i \in [N] : Q_i(t) > 0 \}$ denote the set of users with non-empty queues. The policy then samples the state of the machine with probability $\mu(\mathcal{S}(t))$ and schedules a user according to the probability distribution $\bm{\pi}(\mathcal{S}(t))$, that is, it selects user $i$ with probability $\pi_i(\mathcal{S}(t))$.
\end{definition}

\begin{remark}
    An adaptive stationary randomized policy can be viewed as a collection of pairs of stationary randomized policies, one governing user selection and one governing sampling, each supported on a distinct subset of $[N]$.    
\end{remark}

In a \textit{conventional stationary randomized scheduling policy}, the central server schedules a user with a fixed probability throughout the entire time horizon, e.g., see~\cite{kadota18}. However, at a given time slot, a job queue can be empty; thus, scheduling that user during that slot with a non-zero probability would waste system resources. Thus, using system resources efficiently while being mathematically tractable motivates the proposed adaptive randomized policy. Similar to the definition of an adaptive randomized policy in Definition~\ref{def:1}, we formally define an adaptive randomized sampling policy. 
\begin{definition}
    An adaptive randomized sampling policy is specified by the collection
    \begin{align}
        \bar{\Pi} = \{\, \mu(\mathcal{S})  : \mathcal{S} \subseteq [N],\, \mathcal{S} \neq \varnothing \,\},
    \end{align}
    where  for each non-empty subset $\mathcal{S} \subseteq [N]$, $\mu(\mathcal{S}) \in (0,1]$ is the sampling probability. At each time slot $t$, let $\mathcal{S}(t) = \{ i \in [N] : Q_i(t) > 0 \}$ denote the set of users with non-empty queues. The policy then samples the state of the machine with probability $\mu(\mathcal{S}(t))$.
\end{definition}

At time $t$, the state of the system is described by the vector $\bm{y}(t) = \left( (Q_i(t))_{i=1}^N, (v_i(t))_{i=1}^N, \: x(t), \: \gamma(t)\right)$, where $\gamma(t)$ represents the number of slots since the server last knew the true state of the machine. We denote the set of all possible states by $\mathcal{Y}$. Thus,  
\begin{align}
    \mathcal{Y} =  \mathbb{N}^{N}\times\mathbb{N}^{N}\times \{-1,0,\cdots,N,f_{r}\}\times \mathbb{N}.
\end{align}
Note that under the proposed policies, $\bm{y}(t)$ evolves as a Markov chain. 
Next, we define the notion of stability~\cite{tassiulas1990stability}.

\begin{definition} \label{def:2}
The system is stable if all recurrent states of the Markov chain $(\bm{y}(t))_{t=1}^\infty$ are positive recurrent and the process hits the recurrent states with probability one. Furthermore, when the whole system is stable we say that the queues are also stable.
\end{definition}

\section{Results}
\subsection{Adaptive Randomized Policy}\label{subsec:adap}
We consider a non-empty subset of users, denoted by $\mathcal{S}$, from the set of $N$ users.  We assume that the job queues corresponding to the users in $\mathcal{S}$ are always non-empty, i.e., at any given time $t$, if the central server receives information that the machine is in the free state, it can assign a job to the machine from a job queue corresponding to one of the users from the set $\mathcal{S}$. In Theorem~\ref{th:1}, we find the total average age of job completion under a conventional stationary randomized policy $\phi=(\mu,\pi)$, corresponding to the users in set $\mathcal{S}$.

\begin{theorem}\label{th:1}
    Consider a fixed subsystem comprising a non-empty subset of users $\mathcal{S} \subseteq [N]$, where each user $i \in \mathcal{S}$ has arrival rate $p_i$ and service rate $q_i$. Let the machine parameters be $q$ and $s$. If the job queues of all users in $\mathcal{S}$ remain non-empty throughout the entire time horizon $T$, then the average age of user $k \in \mathcal{S}$ under a conventional stationary randomized policy $\phi=(\mu, \bm{\pi}) \equiv (\mu(\mathcal{S}), \bm{\pi} (\mathcal{S}))$ is given by
    \begin{align}
        \Delta_k^{\phi}(\mathcal{S}) &\!=\! \frac{1}{\left( \frac{s}{q} \!+\! 2\left( \frac{1}{\mu} \!-\! 1 \right) \!+\! \bar{\eta} \right)} \Bigg( \frac{\psi_k^2}{\pi_k}  
        \!+\! \left( \frac{1}{q_k}\! +\! \frac{1 \!-\! s}{q}\! -\! 2 \right) \psi_k  \notag \\
        &\qquad \qquad + \frac{1}{q} \left( (1 - s)(1 - \pi_k - \eta_k) - \frac{1}{\mu} \right) \notag   \\
        & \qquad \qquad - \frac{\pi_k(1 - q_k)}{q_k} + \sum_{i\in \mathcal{S}} \frac{\pi_i(1 - q_i)}{q_i^2}
        \Bigg) + 1, \label{eq:D1_exp}
    \end{align}
    where  $\bar{\eta} \equiv \bar \eta (\phi)= \sum_{i \in \mathcal{S}} \frac{\pi_i}{q_i}$, $\eta_k \equiv \eta_k(\phi) = \bar{\eta} - \frac{\pi_k}{q_k}$ and $\psi_k \equiv \psi_k(\phi) =  \eta_k + \pi_k + 2\left( \frac{1}{\mu} -1\right) + \frac{s}{q}$.
\end{theorem}

In Theorem~\ref{th:2}, under a similar setting as for Theorem~\ref{th:1}, we find an upper bound for the average sampling cost for a conventional stationary randomized policy $\phi=(\mu, \bm{\pi})$.

\begin{theorem}\label{th:2}
    For a subset of users $\mathcal{S}$ from the set of $N$ users, if the job queues of all users in $\mathcal{S}$ remain non-empty throughout the entire time horizon $T$, then under a stationary randomized policy $\phi=(\mu, \bm{\pi}) \equiv (\mu(\mathcal{S}), \bm{\pi} (\mathcal{S}))$, the average sampling cost $  S^{\phi}(\mathcal{S})$ is upper bounded by $S^{\phi}_{ub}(\mathcal{S})$, where
    \begin{align}
      S^{\phi}_{ub}(\mathcal{S})=\frac{(L+1)\mu}{p^{*}} \left(\frac{1}{\frac{1}{\mu p^{*}}+ \bar\eta }\right),
    \end{align}
    with $p^{*} = \frac{q}{1-(1-2q)(1-\mu)}$.    
\end{theorem}

In both Theorem~\ref{th:1} and Theorem~\ref{th:2}, we consider a fixed subsystem consisting of a subset of users $\mathcal{S}$ whose queues remain permanently backlogged, operating under a conventional stationary randomized policy $\phi=(\mu, \bm{\pi}) \equiv (\mu(\mathcal{S}), \bm{\pi} (\mathcal{S}))$. These theorems, when combined, provide a closed-form expression, in terms of $\mu$ and $\bm{\pi}$, that serves as an upper bound on the total average cost of the subsystem under the considered policy. Looking at the subsystem of $\mathcal{S}$ users, we consider the following optimization problem,
\begin{align}
    &\min_{\phi} \ \sum_{k \in \mathcal{S}} \Delta^{\phi}_{k} (\mathcal{S}) + S^{\phi}_{ub}(\mathcal{S}) 
    \nonumber \\
    &\text{ s.t.} \quad \mu \in (0,1), \quad \pi_k \in (0,1), \: k \in \mathcal{S}, \nonumber \\
    & \quad \quad \sum_{k \in \mathcal{S}}\pi_k =1. \label{eq:opt_prob} 
\end{align}
We numerically solve (\ref{eq:opt_prob}) and obtain a local minimizer denoted by $(\mu^*(\mathcal{S}), \bm{\pi}^*(\mathcal{S}))$.
Now, recalling that an adaptive  randomized policy is a collection of conventional stationary randomized policies, we construct such a collection, denoted by $\Pi_{c}$, by gathering the solutions of the above optimization problem for every non-empty subset of users $\mathcal{S}$, i.e.,
\begin{align}\label{eq:n12}
    \Pi_{c} = \bigcup_{\substack{\mathcal{S} \subseteq[N] \\ S \neq \varnothing}} \{ (\mu^*(\mathcal{S}), \bm{\pi}^*(\mathcal{S})) \}.
\end{align}
We consider an adaptive randomized policy $\phi_{1}$, characterized by $\Pi_{c}$, given in (\ref{eq:n12}). In Section~\ref{sec:num}, we numerically evaluate the performance of $\phi_1$.

\subsection{Max-Age Scheduling Policy with Adaptive Randomized Sampling Policy}
Similar to Sub-section~\ref{subsec:adap}, we consider a non-empty subset of users $\mathcal{S}$, and assume that the job-queues corresponding to the users in $\mathcal{S}$ are always non-empty. Now, we consider a scheduling policy that, at each time slot $t$, schedules the user in the set $\mathcal{S}$ with the maximum age of job completion. We call this the max-age policy and denote it with $\pi^{MA}(\mathcal{S})$. Note that, in our setting and because of the dynamics of age of job completion, $\pi^{MA}(\mathcal{S})$ corresponds to the round-robin policy on set $\mathcal{S}$. Now, we consider a policy $\bar{\phi}$ that consists of a conventional stationary randomized policy $\bar{\mu}$ and the max-age scheduling policy. In the Theorem~\ref{th:4}, we find the total average age of job completion of the users in $\mathcal{S}$, under $\bar{\phi}$.

\begin{theorem}\label{th:4}
    For a subset of users $\mathcal{S}$ from the set of $N$ users, if the job queues of all users in $\mathcal{S}$ remain non-empty throughout the entire time horizon $T$, then under a max-age policy $\bar{\phi}=(\bar{\mu},\pi^{MA})\equiv (\bar{\mu}(\mathcal{S}),\pi^{MA}(\mathcal{S}))$, the average age of user $k$ is given by
    \begin{align}
        \Delta_{k}^{\bar{\phi}}(\mathcal{S}) = &\frac{N \left(\beta_2 - \beta_1^2 \right) + \sum_{i \in \mathcal{S}} \frac{1-q_i}{q_i^2}}{2\left(N\beta_1 + \sum_{i \in \mathcal{S}} \frac{1-q_i}{q_i} \right)} \nonumber \\ 
        &  +\frac{1}{2}\left(N\beta_1 + \sum_{i \in \mathcal{S}} \frac{1-q_i}{q_i} + 1\right),
    \end{align}
    where 
    \begin{align*}
        \beta_1 &= \frac{1}{\bar{\mu}} \left((1-\bar{\mu})+s\frac{\bar{\mu}}{q} +1 \right), \\
        \beta_2 &= 2\frac{1-\bar{\mu}}{\bar{\mu}}\left( \frac{s}{1-\bar{\mu}} \alpha^2 + \alpha -s - \bar{\mu}(1-s) +3\right) + \beta_1,
    \end{align*}
    with $\alpha = 1-\bar{\mu}+\frac{\bar{\mu}}{q}$.        
\end{theorem}

In Theorem~\ref{th:3}, we find the total sampling cost corresponding to the policy $\bar{\phi}$.
\begin{theorem}\label{th:3}
    For a subset of users $\mathcal{S}$ from the set of $N$ users, if the job queues of all users in $\mathcal{S}$ remain non-empty throughout the entire time horizon $T$, then under a policy $\bar{\phi}=(\bar{\mu},\pi^{MA})\equiv (\bar{\mu}(\mathcal{S}),\pi^{MA}(\mathcal{S}))$, the average sampling cost $  S^{\bar{\phi}}(\mathcal{S})$ is given as,
    \begin{align}
      S^{\bar{\phi}}(\mathcal{S})=&\frac{\bar{\mu} N L}{N\left((1-\bar{\mu}) +  \frac{s\bar{\mu}}{q}+1\right) + \bar{\mu}\sum_{i\in\mathcal{S}}\frac{1-q_{i}}{q_{i}}} \nonumber\\
      &\left(p_{1}^{*} \!+\! \frac{(1\!-\!p_{1}^{*})(1\!+\!p^{*})}{p^{*}}\right),
    \end{align}
    where $p_{1}^{*} = s (1-\bar{\mu})p^{*} + (1-s)(1-(1-\bar{\mu})p^{*})$.
\end{theorem}

Now, similar to (\ref{eq:opt_prob}), for a subset of users $\mathcal{S}$,  we consider the following optimization problem,
\begin{align}
     &\min_{\bar{\phi}} \ \sum_{k \in \mathcal{S}} \Delta^{\bar{\phi}}_{k} (\mathcal{S}) + S^{\bar{\phi}}(\mathcal{S}) \nonumber \\
    &\text{ s.t.} \quad \bar{\mu} \in (0,1). \label{eq:opt_prob1}
\end{align}
We numerically solve (\ref{eq:opt_prob1}) and obtain a local minimizer denoted by $\bar{\mu}^{*}(\mathcal{S})$. Now, consider the following collection of conventional stationary randomized sampling policies,
\begin{align}\label{eq:n19}
    \bar{\Pi}_{c} = \bigcup_{\substack{\mathcal{S} \subseteq[N] \\ S \neq \varnothing}} \{ \bar{\mu}^*(\mathcal{S}) \}.
\end{align}
We consider a policy $\bar{\phi}_{1}$, comprising an adaptive randomized sampling policy characterized by $\bar{\Pi}_{c}$ in (\ref{eq:n19}), and the scheduling policy $(\pi^{MA}(\mathcal{S}(t)))_{t=1}^{\infty}$, where $\mathcal{S}(t) = \{ i \in [N] : Q_i(t) > 0 \}$. In Section~\ref{sec:num}, we numerically evaluate its performance.

\subsection{Stabilizing the Queues}
We now derive sufficient conditions for queue stability under both an adaptive randomized policy and a max-age policy. In both cases, the state of the system $\bm{y}(t)$ evolves as a Markov chain and we follow Definition~\ref{def:2} for stability.

\begin{prop} \label{prop:N_users}
    Consider a system with $N$ users having arrival rates $p_i$, service rates $q_i$,  and machine parameters $q$ and $s$. An adaptive randomized policy $\phi$ characterized by
    \begin{align}
        \Pi = \{\, (\mu(S), \bm{\pi}(S)) : S \subseteq [N],\, S \neq \varnothing \,\},
    \end{align}
    stabilizes the system if there exists $\epsilon > 0$ such that, for every non-empty subset $\mathcal{S} \subseteq [N]$,
    \begin{align}
        \sum_{j = 1}^N p_j 
        - \mu(\mathcal{S})\,(1-\chi(q,s))
        \sum_{i \in \mathcal{S}} \pi_i(\mathcal{S})\,q_i 
        \le -\epsilon,
    \end{align}
    where 
    \begin{align*}
        \chi(q, s) = \begin{cases}
        1-q, & \text{if } q \!\in\! (0,\frac{1}{2}], \: s \!\in\! (0,1), \\
        \frac{(1-2q) \min \{ 1-2q, 2s-1\}+1}{2}, & \text{if } q \!\in\! (\frac{1}{2},1), \: s \!\in\! (0,\frac{1}{2}], \\
        \frac{(1-2q)^2(2s-1) + 1}{2}, & \text{if } q \!\in\! (\frac{1}{2},1), \: s \!\in\! (\frac{1}{2},1). \\
    \end{cases}
    \end{align*}
\end{prop}

\begin{remark}    
    Note that $1 - \chi(q,s)$ provides a lower bound on the probability that the machine is free when sampled. With this interpretation, the conditions in Proposition~\ref{prop:N_users} ensure that the total arrival rate does not exceed the worst-case effective service rate under any of the stationary randomized policies. Conditions of this nature are, in general, known to be necessary and sufficient for rate stability \cite{neely2010stochastic}. Here, we further show that they stabilize the queues according to the notion of stability in Definition~\ref{def:2}, under the nontrivial dynamics of the considered model.
\end{remark}

Proposition~\ref{prop:N_users} asks for an exponential number of conditions to hold for the queues to be stable. A single, but more restrictive, sufficient condition for stability is given in Corollary~\ref{cor:1}.

\begin{corollary}\label{cor:1}
    Under any adaptive randomized policy, the system is stable if there exists $\epsilon >0$, such that
    \begin{align}
        \sum_{i=1}^N p_i - \mu_{\text{min}}(1-\chi(q,s))q_{\text{min}} \leq - \epsilon,
    \end{align}
    where $\mu_{\text{min}} \!=\! \min_{\mathcal{S} \subseteq [N], S \neq \varnothing} \{ \mu(S)\}$ and $q_{\text{min}} \!=\! \min_{i \in [N]}\{q_i\}$.
\end{corollary}

To obtain sufficient conditions for queue stability under the max-age policy, we follow a similar line of reasoning leading to the statement in Proposition~\ref{prop:max_weight}

\begin{prop} \label{prop:max_weight}
    Consider a system with $N$ users having arrival rates $p_i$, service rates $q_i$,  and machine parameters $q$ and $s$. A max-age policy $\bar{\phi} = \left(\mu, \pi^{MA}\right)$ stabilizes the system if there exists $\epsilon > 0$ such that for every non-empty subset $\mathcal{S} \subseteq [N]$, 
    \begin{align}
        \sum_{j = 1}^N p_j - \mu(\mathcal{S})(1-\chi(q,s))q_{\text{min}}(\mathcal{S}) \le -\epsilon,
    \end{align}
    where $q_{\text{min}}(\mathcal{S}) = \min_{i \in \mathcal{S}} \{ q_i\}$.
\end{prop}

\section{Numerical Results}\label{sec:num}
\begin{figure}
    \centering
    \includegraphics[width=0.85\linewidth]{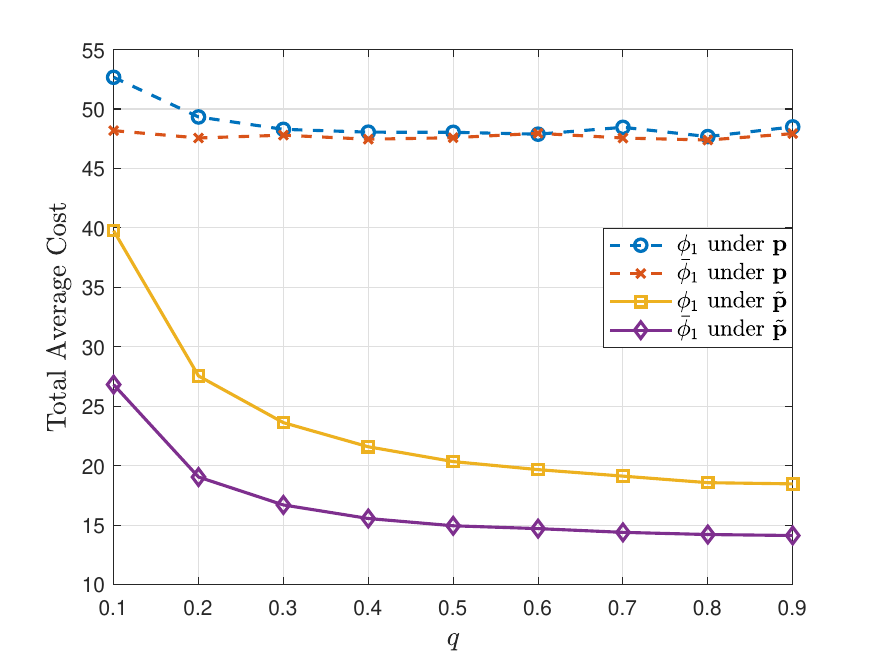}
    \caption{Total average cost  of policies $\phi_1$ and $\bar{\phi}_1$ as a function of $q$, under two distinct arrival rate configurations, $\bm{p}$ and $\tilde{\bm{p}}$, for a system of $N=4$ users.}
    \label{fig:plot}
    \vspace*{-0.4cm}
\end{figure}

We begin by comparing the performance of the adaptive randomized policy, denoted by $\phi_1$, against the max-age scheduling policy with adaptive randomized sampling, denoted by $\bar{\phi}_1$. Fig.~\ref{fig:plot} shows the total cost incurred by both policies as a function of $q$, evaluated under two distinct arrival rate configurations. Specifically, we consider a system with $N=4$ users, $s = 0.5$, sampling cost $L=5$ and service rates given by the vector $\bar{\bm{q}} = [0.1 \; 0.4 \; 0.6 \; 0.9]$. The two arrival rate configurations are given by the vectors $\bm{p} = [0.01 \; 0.02 \; 0.05 \; 0.06]$  and $\tilde{\bm{p}} =[0.05 \; 0.2 \; 0.5 \; 0.6]$. In Fig.~\ref{fig:plot} we see that the max-age policy $\bar{\phi}_1$ consistently outperforms the adaptive randomized $\phi_1$, with the performance gap being significant under the high arrival rate configuration $\tilde{\bm{{p}}}$. Under the same arrival rate configuration, the total cost decreases with $q$ for both policies. Moreover, under the low arrival rate configuration $\bm{p}$,  both policies perform similarly, which is expected since both policies are work-conserving and the arrival rates are low. 

We observe numerically that queue stability is achieved only when the users' arrival rates are well below their respective service rates. For example, with model parameters $N=4$, $q = 0.35$, $s = 0.3$, $L=5$,  service rates $\bar{\bm{q}} = [0.55 \; 0.73 \; 0.84 \; 0.91]$, and arrival rates $\bm{p} = [0.09 \; 0.09 \; 0.12 \; 0.14]$, the conditions in Propositions~\ref{prop:N_users} and~\ref{prop:max_weight} are not satisfied and the queues are found numerically to not be stable.
Furthermore, the conditions stated in Propositions~\ref{prop:N_users} and~\ref{prop:max_weight} are observed to be only sufficient and not necessary for queue stability. For model parameters $N=4$, $q=0.5$, $s = 0.5$, $L=5$, service rates $\bar{\bm{q}} = [0.4 \; 0.6 \; 0.8 \; 0.94]$ and arrival rates $\bm{p} = [0.04 \; 0.05 \; 0.06 \; 0.06]$, the sufficient stability conditions are not satisfied; however,  numerical results show that the queues remain stable both under $\phi_1$ and $\bar\phi_1$. 

\bibliographystyle{unsrt}
\bibliography{references}
\end{document}